\documentclass[12pt]{article}
\usepackage{graphics}
\begin{document}
\begin{center}{
{\Large \bf Quantum fidelity for Gaussian states \\
describing the evolution of open systems} \vskip 0.5truecm
Aurelian Isar \\
{National Institute of Physics and Nuclear Engineering\\
Bucharest-Magurele, Romania }\\
{\it isar@theory.nipne.ro}
}
\end{center}

\abstract{Using the expression of the fidelity for the most general
Gaussian quantum states, the quantum fidelity is
studied for the states of a harmonic oscillator interacting with
an environment, in particular with a thermal bath. The time evolution
of the considered system is described in the framework of the
theory of open systems based on quantum dynamical semigroups. By taking a correlated squeezed Gaussian state as initial state, we calculate
the quantum fidelity for both undisplaced and displaced
states. The time evolution of the quantum fidelity is analyzed depending on the squeezing and correlation parameters characterizing the initial Gaussian state and on the dissipation constant and temperature of the thermal bath.
} 
%
%
\section{Introduction}
In quantum information theory it is very important to
quantify the similarity or distinguishability of quantum states.
There are several distance measures widely used to distinguish
quantum states in quantum information transfer, in particular the
trace distance and quantum fidelity \cite{nie}. Using Uhlmann's
transition probability \cite{uhl}, Jozsa \cite{joz} defined the
quantum fidelity of two quantum states. The fidelity quantifies how
close are the two states, and their separation can be measured, for
example, by the Bures distance \cite{bur}.

In the recent years there has been an increased interest in the
study of Gaussian states of continuous variable systems used in
quantum information processing \cite{bra}. Usually the quantum
fidelity for quantum optics experiments, in particular quantum
teleportation experiments, is calculated for pure coherent states as
input states \cite{bow,zha,lan}. However, in real experiments the
input quantum states have some non-negligible degree of mixedness,
achieved mainly due to the decoherence phenomenon which takes place
during the interaction of the system with its environment. We mention here just a few of the many studies on quantum fidelity of Gaussian states. In Ref. \cite{jeo} the authors made intense investigations in order to understand the effects of the mixedness of states on the quantum fidelity achieved by experiments and the significance of the obtained results for the fidelity of Gaussian states. The quantum fidelity of Gaussian channels is studied in Refs. \cite{cav,qin} and the fidelity for quantum information protocols is investigated in Refs. \cite{jeo,cav,qin}, in particular for quantum cloning and continuous-variable teleportation.

In the present paper we consider the simple system consisting of an one-dimensional
harmonic oscillator interacting with an environment, in particular with a thermal bath. Our purpose is to study, in the framework of the theory of open systems based on completely positive quantum dynamical semigroups, the influence of the environment on the time evolution of the quantum fidelity for
Gaussian states of the considered system. The structure of the paper is the following. In Sec. 2 we briefly
review the basic formalism for the calculation of the quantum
fidelity for a general pair of single-mode Gaussian states. The time
evolution of the harmonic oscillator in the theory of open quantum
systems is described in Sec. 3. By taking a correlated squeezed
Gaussian state as initial state, the quantum fidelity of this initial pure
state and an arbitrary time mixed state of the considered system is
calculated in Sec. 4. Conclusions are given in Sec. 5.

\section{Fidelity of Gaussian states}
%
For two quantum states, described by the density operators $\rho_1$
and $\rho_2,$  the fidelity $F$ is defined as the Uhlmann's
transition probability by \cite{uhl}
\begin{equation}F(\rho_1,\rho_2)=\left[{\rm Tr}
(\mkern-3mu\sqrt{\rho_1}\mkern3mu\rho_2\sqrt{\rho_1})^\frac{1}{2}
\right]^2.
\end{equation}
The fidelity $F$ is continuous with respect to $\rho_1$ and $\rho_2$
and it has the following natural properties as a measure of the
closeness of the two states:
\par
\begin{itemize}
\item[(i)]
$0\le F(\rho_1,\rho_2)\le 1$ and $F(\rho_1,\rho_2)=1$ if and only if
$\rho_1=\rho_2$;
\item[(ii)]
$F(\rho_1,\rho_2)=F(\rho_2,\rho_1)$ (symmetry);
\item[(iii)]
$F(\rho_1,\rho_2)=\langle\psi_1|\rho_2|\psi_1\rangle$ if one of the
states is pure, {\it i.e.}, when $\rho_1=|\psi_1\rangle\langle
\psi_1|$;
\item[(iv)]
$F(U\rho_1 U^\dag,U\rho_2 U^\dag)=F(\rho_1,\rho_2), $ {\it i.e.}, $F(\rho_1,\rho_2)$ is invariant under a unitary transformation $U$
on the state space.
\end{itemize}

Gaussian states have positive definite Wigner functions which may be
interpreted as probability distributions in phase space. A general
single-mode Gaussian state $\rho$ is completely characterized by its
first and second moments and can be represented by a correlated
squeezed state \cite{dodkur} or by a squeezed and displaced thermal
state \cite{mar}.

The fidelity of two undisplaced thermal states was calculated in
Ref. \cite{twa} and that of displaced thermal states in Ref.
\cite{par}. The fidelity of a large class of single-mode Gaussian
states was obtained in Ref. \cite{scu}, using the fact that the corresponding density operators belong to the oscillator semigroup.

The technical difficulty in calculating the fidelity comes from the
square root of operators. For Gaussian states this difficulty is
solved, since the characteristic function of the square root of a
Gaussian state is also Gaussian. By a successive use of the
composition rule in the position representation,
\begin{eqnarray}\langle x |\rho_1\rho_2|y\rangle=\int_{-\infty}^{+\infty} dz\langle
x|\rho_1|z\rangle\langle z|\rho_2|y\rangle\end{eqnarray} with
\begin{eqnarray}\langle
x|\rho|y\rangle=\exp[-(ax^2+dy^2+2bxy)+lx+ky+g],\label{gaus}
\end{eqnarray} it was obtained in Ref. \cite{scu} the quantum fidelity of an undisplaced and unsqueezed thermal state $\rho_1$ and a displaced squeezed
state $\rho_2.$ In Eq. (\ref{gaus}) the parameters $a, b, d, k, l$ and $g$ have to fulfil the corresponding constraints imposed by the fact that the quantum density operator $\rho$ is Hermitean, normalizable and non-negative. Denoting the amplitude of $\rho_2$ by the row vector
${\bf \alpha}\equiv \left(\matrix{\alpha_{q}\cr\alpha_{p}\cr}\right)
$ and introducing the matrices ${\bf A}_i$ $(i=1,2)$ of the form
\begin{equation}
{\bf A}= \left(\matrix{a_{qq}&a_{pq}\cr
a_{pq}&a_{pp}\cr}\right),\label{matra}
\end{equation}
whose elements are connected with the variances $\sigma_{qq},$
$\sigma_{pp}$ and covariance $\sigma_{pq}$ of canonical position $q$
and momentum $p$ operators through the relations
$a_{qq}=2\sigma_{qq},$ $a_{pp}=2\sigma_{pp}/\hbar^2,$
$a_{pq}=2\sigma_{pq}/\hbar, $ we have \cite{scu}
\begin{eqnarray}
F=\frac{2}{\sqrt{\Delta+\delta}-\sqrt{\delta}}\mkern3mu
\exp\mkern-2mu\left[-{\bf\alpha}^{\rm T}({\bf A}_1+{\bf A}_2)^{-1}{\bf
\alpha}\right], \label{fidelity1}
\end{eqnarray}
where
\begin{eqnarray}
\Delta=\det({\bf A}_1+{\bf A}_2),~~\delta= (\det{\bf A}_1-1)(\det{\bf
A}_2-1).
\end{eqnarray}

Using the property (iv) and denoting the mean amplitudes of $\rho_1$
and respectively $\rho_2$ by ${\bf
\alpha}_i\equiv\left(\matrix{\alpha_{qi}\cr\alpha_{pi}\cr}\right)$
($i=1,2$), we obtain a similar formula for the quantum fidelity of
two general Gaussian quantum states:
\begin{eqnarray}
F=\frac{2}{\sqrt{\Delta+\delta}-\sqrt{\delta}}\mkern3mu
\exp\mkern-2mu\left[-{\bf\beta}^{\rm T}({\bf A}_1+{\bf A}_2)^{-1}{\bf
\beta}\right], \label{fidelity2}
\end{eqnarray}
where now ${\bf \beta}={\bf \alpha}_2-{\bf \alpha}_1.$

If $\rho_1$ is a pure state, then $\det{\bf A}_1=1$ and the fidelity (\ref{fidelity2}) becomes
\begin{eqnarray}
F=\frac{1}{\sqrt{\det\displaystyle\frac{{\bf A}_1+{\bf A}_2}{2}}}\mkern3mu
\exp\mkern-2mu\left[-{\bf\beta}^{\rm T}({\bf A}_1+{\bf A}_2)^{-1}{\bf
\beta}\right].\label{fidelity3}
\end{eqnarray}

\section{Master equation for the harmonic oscillator}
%
In the axiomatic formalism of introducing dissipation in quantum
mechanics, based on quantum dynamical semigroups, the irreversible
time evolution of the open system is described by the following
general quantum Markovian master equation for the density operator
$\rho(t)$ \cite{l1}:
\begin{eqnarray}{d \rho(t)\over
dt}=-{i\over\hbar}[  H, \rho(t)]+{1\over 2\hbar} \sum_{j}([ V_{j}
\rho(t), V_{j}^\dagger ]+[ V_{j}, \rho(t) V_{j}^\dagger
]).\label{lineq}\end{eqnarray} Here $  H$ is the Hamiltonian
operator of the system and $  V_{j},$ $ V_{j}^\dagger $ are
operators on the Hilbert space of the Hamiltonian, which model the
environment. In the case of an exactly solvable model for the damped
harmonic oscillator, the two possible operators $ V_{1}$ and $
V_{2}$ are taken as linear polynomials in coordinate $q$ and
momentum $  p$ \cite{ss,rev} and the harmonic oscillator Hamiltonian
$  H$ is chosen of the most general quadratic form
\begin{eqnarray}
  H=  H_{0}+{\mu \over 2}(  q  p+  p  q),
~~H_{0}={1\over 2m}   p^2+{m\omega^2\over 2}  q^2. \label{ham}
\end{eqnarray} With these choices the master equation (\ref{lineq})
takes the following form: \begin{eqnarray} {d \rho \over
dt}=-{i\over \hbar}[ H_{0}, \rho]- {i\over 2\hbar}(\lambda +\mu) [
q, \rho p+ p \rho]+{i\over 2\hbar}(\lambda -\mu)[  p,
 \rho   q+  q \rho]\nonumber\\
  -{D_{pp}\over {\hbar}^2}[  q,[  q, \rho]]-{D_{qq}\over {\hbar}^2}
[  p,[  p, \rho]]+{D_{pq}\over {\hbar}^2}([  q,[  p, \rho]]+ [ p,[
q, \rho]]). ~~~~\label{mast}
\end{eqnarray} The quantum diffusion coefficients
$D_{pp},~D_{qq},~D_{pq}$ and the dissipation constant $\lambda$
satisfy the following fundamental constraints: $  D_{pp}>0,
D_{qq}>0$ and
\begin{eqnarray} D_{pp}D_{qq}-D_{pq}^2\ge {{\lambda}^2{\hbar}^2\over
4}. \label{ineq}
\end{eqnarray}  In the particular case when the asymptotic state is a Gibbs state $   \rho_G(\infty)=e^{-{  H_0\over kT}}/ {\rm Tr}e^{-{
H_0\over kT}}, $ these coefficients have the form \cite{ss,rev}
\begin{eqnarray} D_{pp}={\lambda+\mu\over 2}\hbar
m\omega\coth{\hbar\omega\over 2kT},~~D_{qq}={\lambda-\mu\over
2}{\hbar\over m\omega}\coth{\hbar\omega\over 2kT},~~D_{pq}=0,
\label{coegib}
\end{eqnarray} where $T$ is the temperature of the thermal bath. In
this case, the fundamental constraints are satisfied only if
$\lambda>\mu$ and
\begin{eqnarray} (\lambda^2-\mu^2)\coth^2{\hbar\omega\over 2kT}
\ge\lambda^2.\label{cons}\end{eqnarray}

The necessary and sufficient condition for translational invariance
is $\lambda=\mu$ \cite{ss,rev}. In the following general values
$\lambda\neq \mu$ will be considered. In this way we violate
translational invariance, but we keep the canonical thermal
equilibrium state.

From the master equation (\ref{mast}) we obtain the following
equations of motion for the expectation values $\sigma_{q}$ and
$\sigma_{p}$ of coordinate and momentum:
\begin{eqnarray}{d\sigma_{q}(t)\over
dt}=-(\lambda-\mu)\sigma_{q}(t)+{1\over m}\sigma_{p} (t),
\label{eqmo1}\end{eqnarray} \begin{eqnarray}{d\sigma_{p}(t)\over
dt}=-m\omega^2\sigma_{q}(t)-(\lambda+\mu)\sigma_{p}(t).
\label{eqmo2}\end{eqnarray} In the underdamped case $(\omega>\mu)$
considered in this paper, with the notation
$\Omega^2\equiv\omega^2-\mu^2$, we obtain \cite{ss,rev}:
\begin{eqnarray}\sigma_q(t)=e^{-\lambda t}\left[(\cos\Omega
t+{\mu\over\Omega}\sin\Omega t) \sigma_q(0)+{1\over
m\Omega}\sin\Omega t\sigma_p(0)\right], \label{sol1}\end{eqnarray}
\begin{eqnarray}\sigma_p(t)=e^{-\lambda t}\left[-{m\omega^2\over\Omega}
\sin\Omega t\sigma_q(0)+ (\cos\Omega t-{\mu\over\Omega}\sin\Omega
t)\sigma_p(0)\right] \label{sol2}\end{eqnarray} and
$\sigma_q(\infty)=\sigma_p(\infty)=0.$

From the master equation (\ref{mast}) we can also obtain the
equations of motion for the variances $\sigma_{qq}, $~$\sigma_{pp}$
and covariance $\sigma_{pq}$ of coordinate and momentum:
\begin{eqnarray}{d\sigma_{qq}(t)\over dt}=-2(\lambda-\mu)\sigma_{qq}(t)+{2\over
m} \sigma_{pq}(t)+2D_{qq},\end{eqnarray}
\begin{eqnarray}{d\sigma_{pp}\over
dt}=-2(\lambda+\mu)\sigma_{pp}(t)-2m\omega^2
\sigma_{pq}(t)+2D_{pp},\end{eqnarray}
\begin{eqnarray}{d\sigma_{pq}(t)\over
dt}=-m\omega^2\sigma_{qq}(t)+{1\over m}\sigma_{pp}(t)
-2\lambda\sigma_{pq}(t)+2D_{pq}.\end{eqnarray} Introducing the
notations
\begin{eqnarray} X(t)=\left(\matrix{m\omega\sigma_{qq}(t)\cr
\displaystyle\frac{\sigma_{pp}(t)}{m\omega}\cr \sigma_{pq}(t)\cr}\right),~~\label{var}
D=\left(\matrix{2m\omega D_{qq}\cr 2\displaystyle\frac{D_{pp}}{m\omega}\cr
2D_{pq}\cr}\right),\end{eqnarray} the solutions of these equations
can be written in the form \cite{ss,rev}
\begin{eqnarray}
X(t)=(Te^{Kt}T)(X(0)-X(\infty))+X(\infty),\label{xvar}\end{eqnarray} where the
matrices $T$ and $K$ are given by
\begin{eqnarray} T={1\over
2i\Omega}\left(\matrix{\mu+i\Omega&{\mu-i\Omega}&{2\omega}\cr
{\mu-i\Omega}&{\mu+i\Omega}&{2\omega}\cr
{-\omega}&{-\omega}&{-2\mu}\cr}\right),~~
K=\left(\matrix{-2(\lambda-i\Omega)&{0}&{0}\cr
{0}&{-2(\lambda+i\Omega)}&{0}\cr
{0}&{0}&{-2\lambda}}\right)\label{matr}\end{eqnarray} and
\begin{eqnarray} X(\infty)=-(TK^{-1}T)D.\label{xinf}\end{eqnarray} Formula
(\ref{xinf}) gives the connection between the asymptotic values
$(t \to \infty)$ of $\sigma_{qq}(t),$ $\sigma_{pp}(t),$ $\sigma_{pq}(t)$
and the diffusion coefficients. These asymptotic values do not depend
on the initial values $\sigma_{qq}(0),$ $ \sigma_{pp}(0),$
$\sigma_{pq}(0)$ and in the case of a thermal bath with coefficients
(\ref{coegib}), they reduce to \cite{ss,rev} \begin{eqnarray}
\sigma_{qq}(\infty)={\hbar\over 2m\omega}\coth{\hbar\omega\over
2kT},~~\sigma_{pp}(\infty)={\hbar m\omega\over
2}\coth{\hbar\omega\over 2kT},~~\sigma_{pq}(\infty)=0. \label{varinf}
\end{eqnarray} In the considered underdamped case we have
\begin{eqnarray} Te^{Kt}T=-{e^{-2\lambda t}\over 2\Omega^2}
\left(\matrix{b_{11}&b_{12}&b_{13}\cr {b_{21}}&{b_{22}}&{b_{23}}\cr
{b_{31}}&{b_{32}}&{b_{33}}\cr}\right),\label{bvar}\end{eqnarray} where $b_{ij}$
$(i,j=1,2,3)$ are time-dependent oscillating functions given by Eqs. (3.78)
in \cite{ss}.

The relation (\ref{ineq}) is a necessary condition for the
generalized uncertainty inequality \begin{eqnarray}\sigma(t)\equiv
\sigma_{qq}(t)\sigma_{pp}(t)-\sigma_{pq}^2(t)\ge{\hbar^2 \over
4}\label{genun1}\end{eqnarray} to be fulfilled, where $\sigma(t)$ is
the Schr\"odinger generalized uncertainty function. The equality in
relation (\ref{genun1}) is realized for a special class of pure
states, called correlated coherent states \cite{dodkur} or squeezed
coherent states.

\section{Calculation of quantum fidelity}
%
We consider a harmonic oscillator with the initial Gaussian wave
function
\begin{eqnarray} \Psi(q)=\left({1\over 2\pi\sigma_{qq}(0)}\right)^{1\over
4}\exp\left[-{1\over 4\sigma_{qq}(0)}
\left(1-{2i\over\hbar}\sigma_{pq}(0)\right)(q-\sigma_q(0))^2 +{i\over
\hbar}\sigma_p(0)q\right], \label{ccs}\end{eqnarray} where
$\sigma_{qq}(0)$ is the initial spread, $\sigma_{pq}(0)$ the initial
covariance, and $\sigma_q(0)$ and $\sigma_p(0)$ are the initial
averaged position and momentum of the wave packet. The initial state
(\ref{ccs}) represents a correlated coherent state \cite{dodkur}
with the variances and covariance of coordinate and momentum
\begin{eqnarray} \sigma_{qq}(0)={\hbar\delta\over 2m\omega},~
\sigma_{pp}(0)={\hbar m\omega\over 2\delta(1-r^2)},~
\sigma_{pq}(0)={\hbar r\over 2\sqrt{1-r^2}},
\label{inw}\end{eqnarray} where $\delta$ is the squeezing parameter
which measures the spread in the initial Gaussian packet and $r,$
with $|r|<1$ is the correlation coefficient at time $t=0.$ The
initial values (\ref{inw}) correspond to a minimum uncertainty
state, since they satisfy the equality in the generalized uncertainty relation (\ref{genun1}):
\begin{eqnarray} \sigma_{qq}(0)\sigma_{pp}(0)-\sigma_{pq}^2(0)
={\hbar^2\over 4}.\label{gen0}\end{eqnarray} For $\delta=1$ and
$r=0$ the correlated coherent state becomes a Glauber coherent
state.

The finite temperature Schr\"odinger generalized uncertainty
function $\sigma(t)$, calculated in Ref. \cite{unc}, has the
expression
\begin{eqnarray} \sigma(t)={\hbar^2\over 4}\{e^{-4\lambda
t}\left[1-\left(\delta+{1\over\delta(1-r^2)}\right)\coth\epsilon
+\coth^2\epsilon\right]\nonumber\\
+e^{-2\lambda
t}[\left(\delta+{1\over\delta(1-r^2)}-2\coth\epsilon\right)
{\omega^2-\mu^2\cos(2\Omega
t)\over\Omega^2}\nonumber \\
+\left(\delta-{1\over\delta(1-r^2)}\right){\mu \sin(2\Omega
t)\over\Omega}+{2r\mu\omega (1-\cos(2\Omega
t))\over\Omega^2\sqrt{1-r^2}}]\coth\epsilon
+\coth^2\epsilon\},\label{sunc}\end{eqnarray} where we have
introduced the notation
\begin{eqnarray} \epsilon\equiv {\hbar\omega\over 2kT}.\end{eqnarray}

We are interested to study the time evolution of quantum fidelity when
the initial state is pure and the system is in interaction with its
environment. If the initial wave function is Gaussian, then the
density matrix remains Gaussian for all times (with time-dependent
parameters which determine its amplitude and spread) and centered
along the trajectory given by Eqs. (\ref{sol1}) and (\ref{sol2}) representing  the solutions $\sigma_q(t)$ and $\sigma_p(t)$ of the dissipative equations of motion \cite{aphysa}. We take the initial Gaussian wave function (\ref{ccs}) as the pure state 1 with the corresponding matrix ${\bf
A}_1$ and the state at an arbitrary time $t$ described by the
density matrix $\rho(t)$ as the state 2 with the corresponding matrix
${\bf A}_2$. Then we can apply the formula for the quantum fidelity
(\ref{fidelity3}), where the matrices ${\bf A}_i$ $(i=1,2)$ of the form (\ref{matra}) are given by
\begin{equation}
{\bf A_1}= \left(\matrix
{a_{qq}(0)&{a_{pq}(0)}\cr{a_{pq}(0)}&{a_{pp}(0)}\cr} \right),~~ {\bf
A_2}= \left(\matrix{
{a_{qq}(t)}&{a_{pq}(t)}\cr{a_{pq}(t)}&{a_{pp}(t)}\cr}\right).
\end{equation}
We obtain \begin{equation}{\bf A}_1+{\bf A}_2=2\left(\matrix
{\sigma_{qq}(0)+\sigma_{qq}(t)
&\displaystyle{\frac{1}{\hbar}}\left(\sigma_{pq}(0)+\sigma_{pq}(t)\right)
\cr\displaystyle{\frac{1}{\hbar}}\left(\sigma_{pq}(0)+\sigma_{pq}(t)\right)
&\displaystyle{\frac{1}{\hbar^2}}\left(\sigma_{pp}(0)+\sigma_{pp}(t)\right)\cr}
\right)\end{equation} and
\begin{eqnarray}\det{{\bf  A}_1+{\bf A}_2\over 2}
=\frac{1}{\hbar^2}[\sigma(0)+\sigma(t) +\sigma_{qq}(0)\sigma_{pp}(t)
+\sigma_{pp}(0)\sigma_{qq}(t)
-2\sigma_{pq}(0)\sigma_{pq}(t)].\end{eqnarray}

\subsection{Case of undisplaced initial state}

In the case when the initial Gaussian wave function (\ref{ccs}) is
not displaced, one has ${\bf \alpha}_1={\bf \alpha}_2=0$ and the
expression (\ref{fidelity3}) simplifies:
\begin{eqnarray}
F=\frac{1}{\sqrt{\det\displaystyle\frac{{\bf A}_1+{\bf A}_2}{2}}}.\label{fidelity4}
\end{eqnarray}
Then we obtain
\begin{eqnarray}
F(t)=\frac{\hbar}{[\sigma(0)+\sigma(t)
+\sigma_{qq}(0)\sigma_{pp}(t) +\sigma_{pp}(0)\sigma_{qq}(t)
-2\sigma_{pq}(0)\sigma_{pq}(t)]^\frac{1}{2}}.\label{fidelity4bis}
\end{eqnarray}
Using formulas (\ref{xvar}), (\ref{varinf}) and (\ref{bvar}) for calculating the variances of the coordinate and momentum, the expression (\ref{fidelity4bis}) of the quantum fidelity takes the following explicit form when the initial state is a squeezed state ($r=0$):
\begin{eqnarray} F(t)=2\{e^{-4\lambda
t}\left[1-\left(\delta+{1\over\delta}\right)\coth\epsilon+\coth^2\epsilon\right]
+\frac{e^{-2\lambda t}}{2\Omega^2}
[\omega^2\left(2+\delta^2+{1\over\delta^2}-4\coth^2\epsilon\right)
\nonumber\\
+4\mu\Omega\left(\delta-{1\over\delta}\right)\sin(2\Omega
t)\coth\epsilon
+\left[\omega^2\left(2-\delta^2-\frac{1}{\delta^2}\right)-4\mu^2(1- \coth^2\epsilon)\right]\cos(2\Omega t)]\nonumber\\+1+\left(\delta+{1\over\delta}\right)\coth\epsilon+
\coth^2\epsilon\}^{-\frac{1}{2}}.\label{fidel5}\end{eqnarray}
In the particular case $\mu=0,$ expression (\ref{fidel5}) takes a simpler form:
\begin{eqnarray} F(t)=2\{e^{-4\lambda
t}\left[1-\left(\delta+{1\over\delta}\right)\coth\epsilon+\coth^2\epsilon\right]
+\frac{e^{-2\lambda t}}{2}
[2+\delta^2+{1\over\delta^2}-4\coth^2\epsilon\nonumber\\
+\left(2-\delta^2-\frac{1}{\delta^2}\right)
\cos(2\omega t)]+1+\left(\delta+{1\over\delta}\right)\coth\epsilon+
\coth^2\epsilon\}^{-\frac{1}{2}}.\label{fidel5bis}\end{eqnarray}

The time evolution of the quantum fidelity (\ref{fidel5}) as a function on the temperature, dissipation constant and squeezing parameter of the initial Gaussian state is represented in Figs. 1--3. From Fig. 1 we see that the fidelity is decreasing with time and temperature: monotonically, in the case of an initial coherent state ($\delta=1$) and non-monotonically if $\delta\neq 0.$ Fig. 2 represents the dependence on time and squeezing parameter. The oscillating behaviour of the fidelity is manifested mainly in the first part of its time evolution. Fig. 3 represents the exponential decreasing of the quantum fidelity with time and dissipation constant: monotonically, in the case of an initial coherent state and non-monotonically if the initial state is squeezed. The oscillating behaviour of the fidelity manifested in the second case is stronger for smaller values of the dissipation constant. It is interesting to remark that in the special case of an initial coherent state ($\delta=1$) and of zero temperature ($T=0$) of the thermal bath, the quantum fidelity has a constant value in time, $F(t)=1,$ independent of the values of the parameter $\mu$ and dissipation constant $\lambda.$
\begin{figure}
\resizebox{1\columnwidth}{!} {
\includegraphics{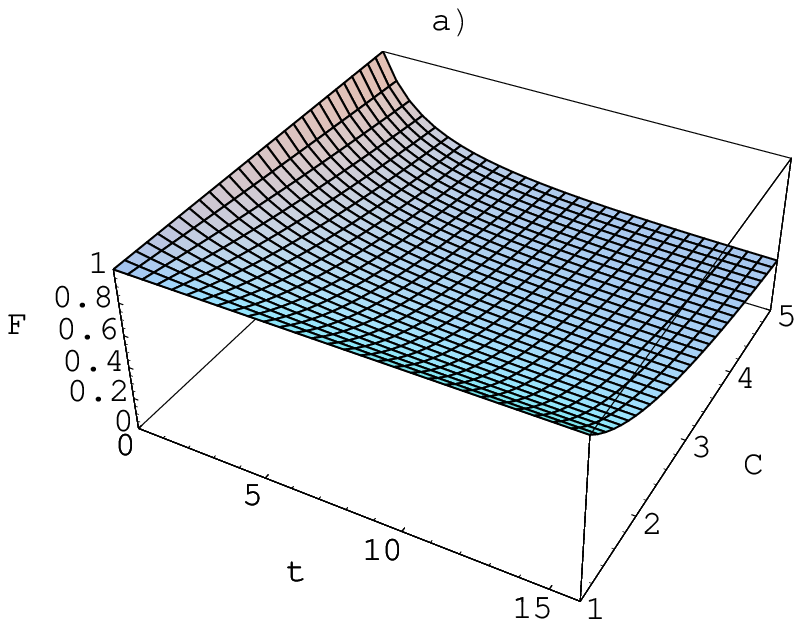}
\includegraphics{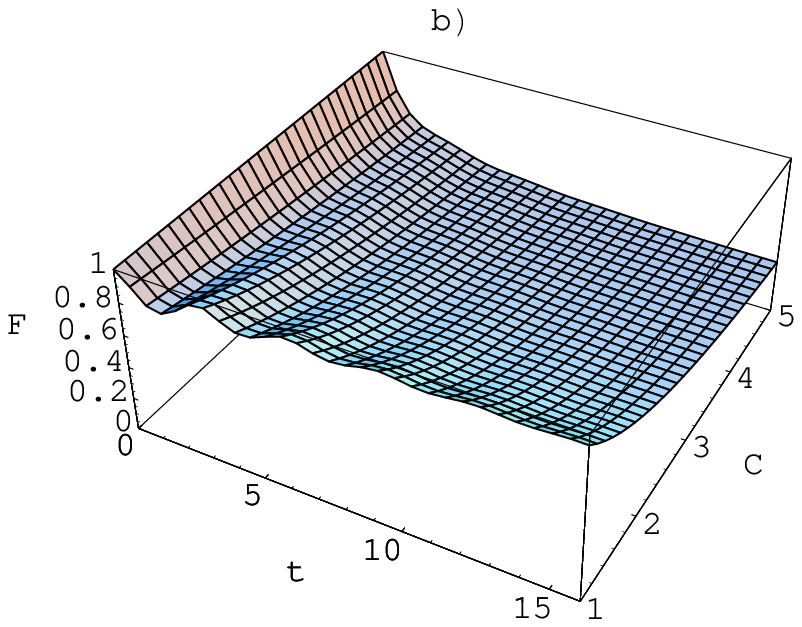}}
\caption{Dependence of quantum fidelity $F$ (\ref{fidel5}) on time
$t\in [0,16]$ and temperature $T$ via  $C\equiv\coth\frac{\hbar\omega}{2kT},$ $C\in [1,5],$ for $\omega=1,$
$\lambda=0.1,$ $\mu=0,$ $r=0.$ Case a) $\delta=1.$ Case b) $\delta=2.$
}
\label{fig:1}       
\end{figure}
\begin{figure}
\resizebox{1\columnwidth}{!} {
\includegraphics{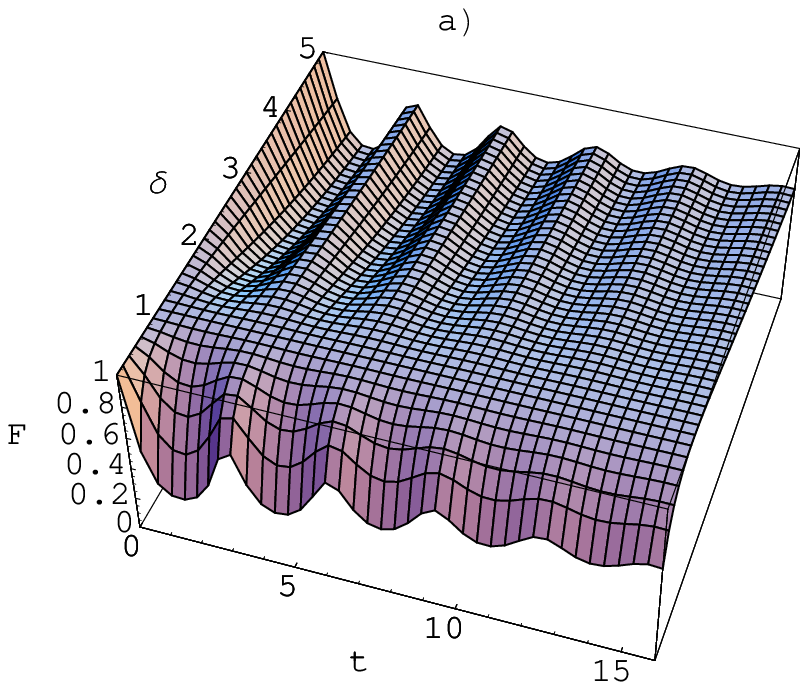}
\includegraphics{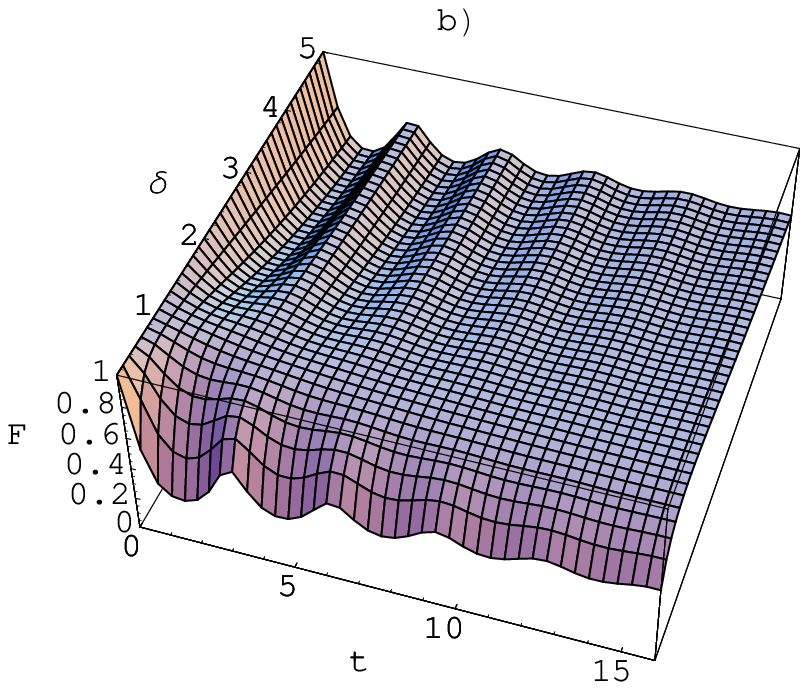}}
\caption{Dependence of quantum fidelity $F$ (\ref{fidel5}) on time
$t\in[0,16]$ and squeezing coefficient $\delta\in [1/8,5],$ for $\omega=1,$
$\lambda=0.1,$ $r=0.$ Case a) $\mu=0,$ $C=1$ $(T=0).$ Case b) $\mu=0.08,$ $C=5/3$ ($C\equiv\coth\frac{\hbar\omega}{2kT}).$
}
\label{fig:2}       
\end{figure}
\begin{figure}
\resizebox{1\columnwidth}{!} {
\includegraphics{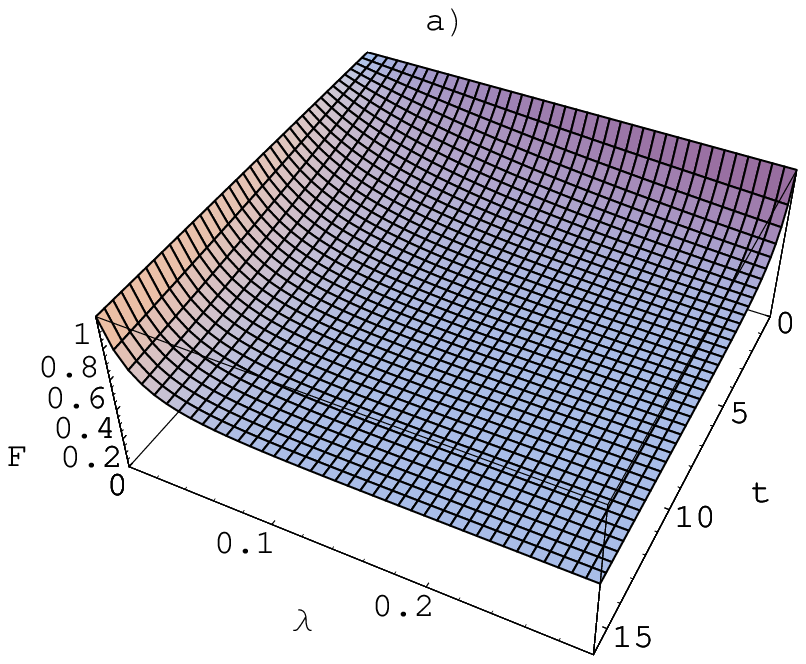}
\includegraphics{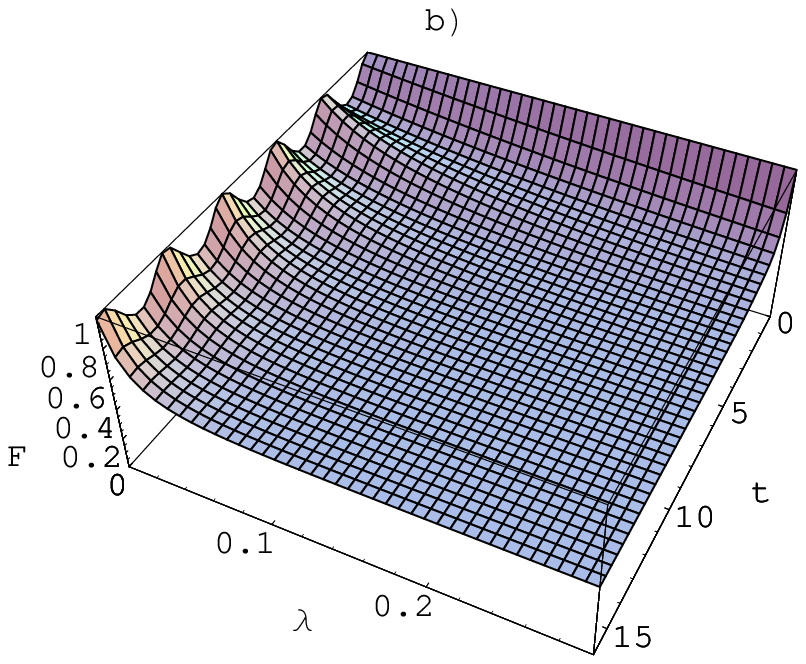}}
\caption{Dependence of quantum fidelity $F$ (\ref{fidel5}) on time
$t\in [0,16]$ and dissipation constant $\lambda\in [0,0.3],$ for $\omega=1,$ $\mu=0,$
$r=0,$  $C=3$ ($C\equiv\coth\frac{\hbar\omega}{2kT}$). Case a)  $\delta=1.$ Case b) $\delta=2.$}
\label{fig:3}       
\end{figure}

The fidelity at the initial moment is evidently always $F(0)=1.$ For
large times we get the following expression of the asymptotic fidelity when the initial state is a correlated squeezed state:
\begin{eqnarray}
F(\infty)=\frac{2}{\sqrt{1+\left(\delta+\displaystyle{\frac{1}{\delta(1-r^2)}}
\right)
\coth\epsilon+\coth^2\epsilon}}.\label{fidelity5}\end{eqnarray}
We notice that the asymptotic value of the quantum fidelity depends on the environment temperature and parameters of the initial state, but does not depend on the dissipation constant.
If $r=0,$ then expression (\ref{fidelity5}) becomes:
\begin{eqnarray}
F(\infty)=\frac{2}{\sqrt{1+\left(\delta+\displaystyle{\frac{1}{\delta}}\right)
\coth\epsilon
+\coth^2\epsilon}}\label{fidelity6}
\end{eqnarray}
and for a coherent state ($\delta=1$) we obtain: \begin{eqnarray}
F(\infty)=\frac{2}{1+\coth\epsilon}.\label{fidelity7}
\end{eqnarray}
When the temperature of the environment is $T=0,$ expression
(\ref{fidelity6}) becomes:
\begin{eqnarray}
F(\infty)=\frac{2\sqrt{\delta}}{\delta+1},\label{fidelity8}
\end{eqnarray}
and respectively $F(\infty)=1,$ if the initial state is a coherent
state. For $\delta\neq 1,$ we have always $F(\infty)<1.$

In Fig. 4 we represent the dependence of the asymptotic value of the quantum fidelity on the environment temperature and squeezing parameter of the initial state. This value is decreasing with increasing temperature. It is also decreasing with the squeezing parameter, when $\delta>1$ and is increasing, when $\delta<1$.
\begin{figure}
\resizebox{1\columnwidth}{!} {
\includegraphics{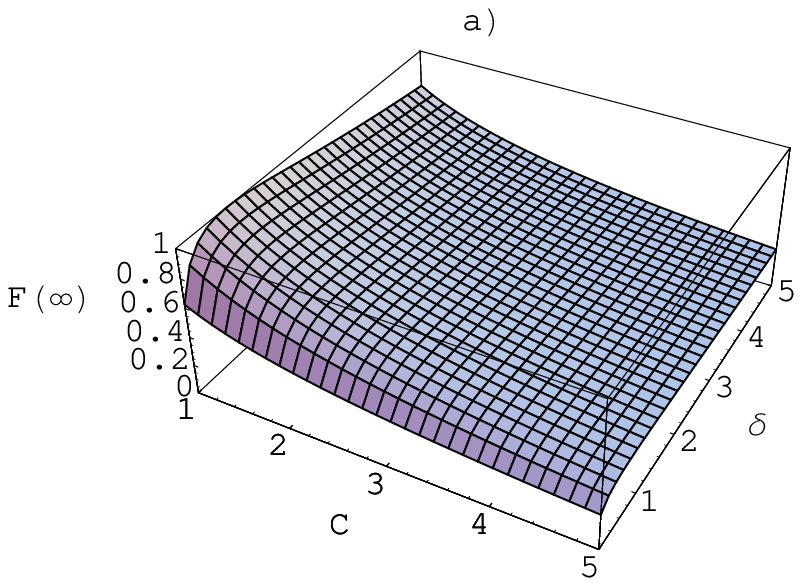}
\includegraphics{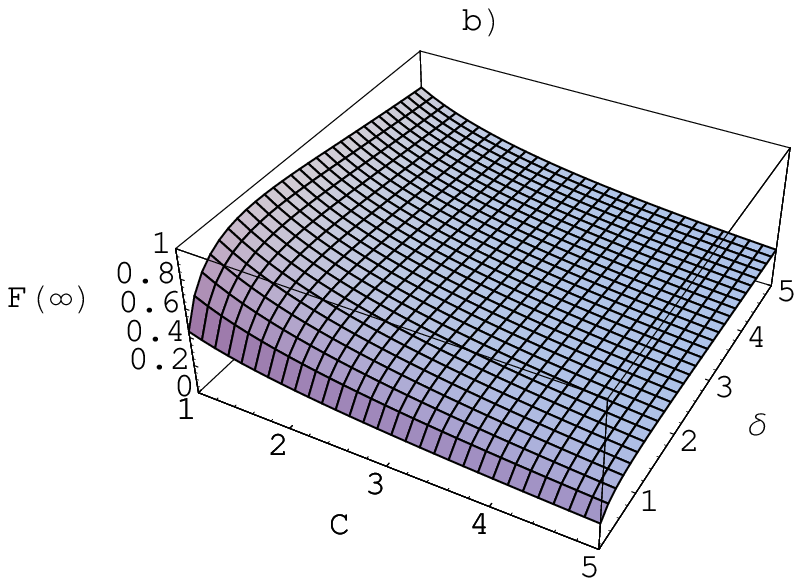}}
\caption{Dependence of asymptotic quantum fidelity $F(\infty)$ (\ref{fidelity5}) on squeezing parameter $\delta \in [1/8,5]$ and temperature $T$ via $C\equiv\coth\frac{\hbar\omega}{2kT},$ $C \in [1,5].$ Case a) $r=0.$ Case b) $r=3/4.$}
\label{fig:4}       
\end{figure}

\subsection{Case of displaced initial state}

In the case of a displaced initial Gaussian wave function (\ref{ccs}), in the previous expressions of the asymptotic fidelity the exponential factor
$\exp(-E)$ of Eq. (\ref{fidelity3}) is present, where $E$ for large
times ($t\to\infty$) has the form
\begin{eqnarray} E(\infty)=\frac{m\omega\sigma^2_q(0)\left(\displaystyle{\frac{1}
{\delta(1-r^2)}}+\coth\epsilon\right)-2
\sigma_q(0)\sigma_p(0)\displaystyle{\frac{r}{\sqrt{1-r^2}}}+\sigma^2_p(0)
\frac{\delta+\coth\epsilon}{m\omega}}{\hbar\left[1+\left(\delta+
\displaystyle{\frac{1}{\delta(1-r^2)}}\right)
\coth\epsilon+\coth^2\epsilon\right]}.\end{eqnarray} For an initial squeezed state ($r=0$) this exponential factor becomes \begin{eqnarray} E(\infty)=
\frac{m^2\omega^2\sigma^2_q(0)\left(\displaystyle{\frac{1}
{\delta}}+\coth\epsilon\right)+\sigma^2_p(0)(\delta+
\coth\epsilon)}{\hbar
m\omega\left[1+\left(\delta+\displaystyle{\frac{1}{\delta}}\right)
\coth\epsilon+\coth^2\epsilon\right]}\end{eqnarray} and for a
coherent state we obtain: \begin{eqnarray} E(\infty)=
\frac{m^2\omega^2\sigma^2_q(0)+\sigma^2_p(0)} {\hbar m\omega
(1+\coth\epsilon)}.\end{eqnarray}
We notice that displacing the initial Gaussian state introduces an additional decreasing exponential factor for the asymptotic quantum fidelity. In the particular case when the temperature
of the environment is $T=0,$ the quantum fidelity for an initial coherent state is given just by the exponential factor:
\begin{eqnarray}
F(\infty)=\exp\left(-\frac{m^2\omega^2\sigma^2_q(0)+\sigma^2_p(0)}
{2\hbar m\omega}\right).\label{fidelity9}
\end{eqnarray}
We see from Eq. (\ref{fidelity9}) that in the case of a displaced coherent state and for a zero temperature of the thermal bath, the asymptotic fidelity is strictly smaller than 1, compared to the case of an undisplaced coherent state, when the asymptotic quantum fidelity is 1. This is due to the fact that the initial mean amplitude of the density operator is decreasing exponentially to zero for asymptotically large times, as a result of dissipation, even in the case of zero temperature of the thermal bath (see Eqs. (\ref{sol1}), (\ref{sol2})).

\section{Conclusions}
%
Using the expression of the fidelity for the most general Gaussian
quantum states, we have studied the quantum fidelity for the states of a harmonic oscillator interacting with a thermal bath. The time evolution
of the considered system was described in the framework of the
theory of open systems based on quantum dynamical semigroups. By
taking a correlated squeezed Gaussian state as the initial state, we
calculated the quantum fidelity for both kinds of undisplaced and
displaced states. We have described the behaviour of the time evolution of the quantum fidelity depending on the squeezing and correlation parameters characterizing the initial Gaussian state and on the dissipation constant and temperature characterizing the environment.

The explicit expressions of the quantum fidelity obtained in the infinite dimensional Hilbert space of density operators can present interest in studying the similarity of Gaussian states of quantum systems with noise and dissipation, that have a time evolution described by completely positive quantum dynamical semigroups. At the same time, the evolution of open quantum systems described by the master equation (\ref{mast}) is equivalent with Gaussian quantum channels with noise and decay. For a noisy quantum channel the state is subject to the decoherence process, which decreases the efficacy of information processing, and a fundamental problem is to determine the fidelity with which the quantum information can be transmitted. The obtained results applied for a bosonic field mode interacting with a thermal reservoir may be therefore useful in the evaluation of the reliability of quantum information processing and communication of Gaussian states through bosonic Gaussian channels. Quantum cloning and continuous-variable teleportation are quantum processes described by bosonic Gaussian channels for which the evaluation of the present results for the quantum fidelity between input and output states might be relevant in measuring the efficiency of these processes.

\section*{Acknowledgments}
%
The author acknowledges the financial support received within
the Projects CEEX-1 68/2005 and PN 06350101/2006.

\end{document}